\documentclass[epjST]{svjour}
\usepackage{graphics}
\begin{document}
\title{Phase Synchronization in Unidirectionally Coupled Ikeda Time-delay
Systems}
\author{D. V. Senthilkumar\inst{1}\fnmsep\thanks{\email{skumar@cnld.bdu.ac.in}}
\and M. Lakshmanan\inst{1}\fnmsep\thanks{\email{lakshmanan@cnld.bdu.ac.in}} 
\and J. Kurths\inst{2}\fnmsep\thanks{\email{jkurths@gmx.de}} }
\institute{Centre for Nonlinear Dynamics, Department of Physics,\\ Bharathidasan
University, Tiruchirapalli - 620 024, India \and Humboldt University,
Berlin and Potsdam Institute for Climate Impact Research, Germany}
\abstract{
Phase synchronization in unidirectionally coupled Ikeda time-delay systems
exhibiting non-phase-coherent  hyperchaotic attractors of complex topology with 
highly interwoven trajectories is studied. It is shown that in this set of
coupled  systems phase synchronization (PS) does exist in a range of the
coupling strength which is preceded by a transition regime (approximate PS) and
a nonsynchronous regime. However, exact generalized synchronization does not seem
to occur in the coupled Ikeda systems (for the range of parameters we have
studied) even for large coupling strength, in contrast to our earlier studies in
coupled piecewise-linear and Mackey-Glass systems~\cite{dvskml2006,dvskml2008}.
The above transitions are characterized  in terms of recurrence based indices,
namely generalized autocorrelation function $P(t)$, correlation of probability
of recurrence (CPR), joint probability of recurrence (JPR) and similarity of
probability of recurrence (SPR). The existence of phase synchronization is also
further confirmed by  typical transitions in the Lyapunov exponents of the
coupled Ikeda time-delay systems and also using the concept of localized sets. }
\maketitle
\section{Introduction}
\label{intro}

Synchronization of chaotic oscillations has been an area of extensive research
since the pioneering works of Fujisaka and Yamada~\cite{hfty1983} and of  Pecora
and Carroll~\cite{lmptlc1990}. Since the identification of complete (identical)
chaotic synchronization,  different kinds of chaotic synchronizations have been
identified and demonstrated both theoretically and
experimentally~(cf.~\cite{lp1997,aspmgr2001,sbjk2002,jk2000}). Among the basic
kinds of synchronization, chaotic phase synchronization (CPS) plays a crucial
role in understanding a large class of weakly interacting nonlinear dynamical
systems in diverse natural systems like cardiac and respiratory systems,
biological clocks synchronized by day and night rhythms, ecological systems
entrained by seasonal cycles, etc.~\cite{aspmgr2001,sbjk2002}. The definition
of  CPS is a direct extension of the classical definition of synchronization of
periodic oscillations and can be referred to as entrainment between the phases
of interacting chaotic oscillators, while their amplitudes remain chaotic and,
in general, uncorrelated~\cite{mgrasp1996,aspmgr1997}.

The notion of CPS has been investigated so far in oscillators driven by
external periodic forces \cite{aspmgr1997,aspgo1997}, chaotic oscillators with
different natural frequencies and/or with parameter mismatches
\cite{mgrasp1996,uplj1996,mzgww2002}, in arrays of coupled chaotic
oscillators \cite{gvoasp1997,mzzz2000} and  also in different
chaotic systems \cite{ercmt2003,sgchl2005}.  In addition CPS has also been
demonstrated experimentally in various systems such as electrical
circuits~\cite{ercmt2003,apoc2003,msbtps2003,skdbb2006},
lasers~\cite{kvvvni2001,djdrb2001}, fluids~\cite{dmav2000}, biological
systems~\cite{ptmgr1998,rceals1998}, climatology~\cite{dmjk2005}, etc.

While the notion of CPS has been well understood in low-dimensional systems  as
mentioned above, it has not yet been studied in detail in nonlinear time-delay
systems. These are essentially infinite-dimensional systems and  correspond to 
an important class of dynamical systems representing several physical phenomena
in diverse areas of science and technology including neuroscience, physiology,
ecology, lasers, etc.,~\cite{dvskthesis2008}. Studying the nature of onset of
CPS and transition to other synchronized states has received considerable
attention recently  due to their importance in understanding the dynamical
nature of the underlying physical systems.  Recently, we have reported the
existence of phase,  CPS and its transition to generalized synchronization (GS)
in coupled time-delay systems such as piece-wise linear and Mackey-Glass
time-delay systems ~\cite{dvskml2006,dvskml2008}, which typically exhibit highly
non-phase-coherent chaotic and hyperchaotic attractors. We have introduced a
nonlinear transformation to capture the phase of non-phase-coherent attractors
of both the systems.  We have also used recurrence based indices such as $P(t)$,
CPR, JPR and SPR directly to the non-phase-coherent  attractors and typical
transitions in the Lyapunov exponents of the coupled time-delay systems to
characterize the synchronization transitions. We have also found that all the
three approaches are in good agreement in indicating the onset of CPS and its
transitions.

In this paper, as a natural extension of our above investigations, we try to
generalize the nonlinear transformation and to test the validity of the above
mentioned recurrence based indices in identifying the synchronization
transitions in general class of time-delay systems  exhibiting highly
non-phase-coherent  hyperchaotic attractors of more complex topology. As an
example, we have considered one of the prototype time-delay systems, namely
Ikeda time-delay system~\cite{kihd1980}, which exhibits highly
non-phase-coherent hyperchaotic attractor with complex topology for suitable
parameter values. Even though, we have not yet succeeded in generalizing the
nonlinear transformation to capture the phase of the non-phase-coherent
hyperchaotic attractors of the Ikeda system,  we found that the recurrence based
indices serve as excellent quantifiers in identifying the transition from
non-synchronized to phase synchronized state both qualitatively and
quantitatively in the coupled Ikeda systems.  We have also  characterized these
transitions by typical changes in the Lyapunov spectrum of the coupled Ikeda
time-delay systems. Further, we have confirmed the existence of CPS using the
concept of localized sets.

The plan of the paper is as follows. In Sec.~II, we briefly point out the
inadequacy of the conventional methods available in the literature in identifying
phase in time-delay systems and the necessity of specialized tools and techniques to
identify phase in such systems, while in Sec.~III we discuss briefly about the
Ikdea time-delay systems and its dynamics. We demonstrate the onset of CPS and
its transition to CPS in coupled Ikeda systems using recurrence based
indices and Lyapunov exponents in Sec.~IV. Finally in Sec.~V, we summarize our
results.

\begin{figure}
\centering
\resizebox{0.75\columnwidth}{!}{
\includegraphics{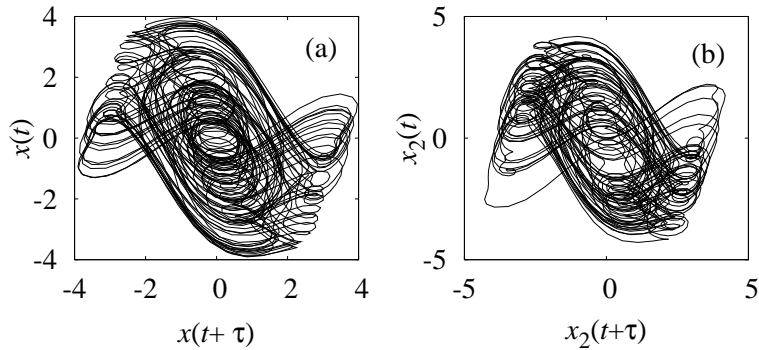}}
\caption{\label{ikeda_att}(a) The non-phase coherent chaotic attractor of the
drive, $x_1(t)$, for the value of delay time $\tau=2$ and (b) the non-phase
coherent hyperchaotic attractor of the response,  $x_2(t)$, for the value of
delay time $\tau=3$ for the Ikeda system~(\ref{eq1.01}).} 
\end{figure}

\section{CPS and Time-delay systems}

As noted in the introduction, CPS has been studied extensively during the last
decade in various nonlinear dynamical systems.  However, only  a few methods are
available in the literature ~\cite{aspmgr2001,sbjk2002} to calculate the phase
of chaotic attractors. Unfortunately most of these measures are restricted to
phase-coherent chaotic attractors, while a few of them are applicable to 
non-phase-coherent chaotic attractors of low-dimensional systems as well. 
However, all these conventional methods which are applicable to the
phase-coherent and non-phase-coherent attractors  cannot be used in the case of
time-delay systems in general, because such systems very often exhibit more
complicated attractors with more than one positive Lyapunov exponents. 
Correspondingly methods to calculate the phase of non-phase-coherent
hyperchaotic attractors of time-delay systems are not available.  The most
promising approach available in the literature to calculate the phase of
non-phase-coherent attractors is based on the concept of
curvature~\cite{gvobh2003}, but this is often restricted to low-dimensional
systems and it does not work in the case of nonlinear time-delay systems in
general, where very often the attractor is non-phase-coherent and
high-dimensional.  This is essentially be due to the  multiple intrinsic
characteristic time scales of the nonlinear time-delay systems. Hence defining
and estimating phase from the hyperchaotic attractors of the time-delay systems
itself is a challenging task and so specialized techniques and  tools have to be
identified to introduce the notion of phase in such systems.

Recently, the present authors have studied in some detail the existence of phase
and CPS in time delay systems admitting non-phase-coherent hyperchoatic
attractors and specifically analyzed coupled piecewise linear and couped
Mackey-Glass systems.  Three different approaches were introduced to identify
and calculate phase and consequently CPS between the interacting time-delay
systems: 1) identifying suitable nonlinear transformation which can unfold the
complicated chaotic and hyperchaotic attractors with multiple loops into smeared
limit cycle like attractor, 2) using recurrence based indices such as
generalized autocorrelation function $P(t)$, correlation of probability of
recurrence (CPR), joint probability of recurrence (JPR) and similarity of
probability of recurrence (SPR) directly to the non-phase-coherent chaotic and
hyperchaotic attractors, we have demonstrated the existence of CPS both
qualitatively and quantitatively and 3) finally the onset of CPS and their
transition is also characterized by typical transitions in the spectrum of
Lyapunov exponents of the coupled time-delay systems. We now apply these
approaches to coupled Ikeda systems which exhibit even more complicated chaotic
and hyperchaotic non-phase-coherent attractors with complex topological
properties having highly interwoven trajectories and identify the nature of CPS.
We have also confirmed the existence of CPS using the framework of localized
sets.

\begin{figure}
\centering
\resizebox{0.8\columnwidth}{!}{
\includegraphics{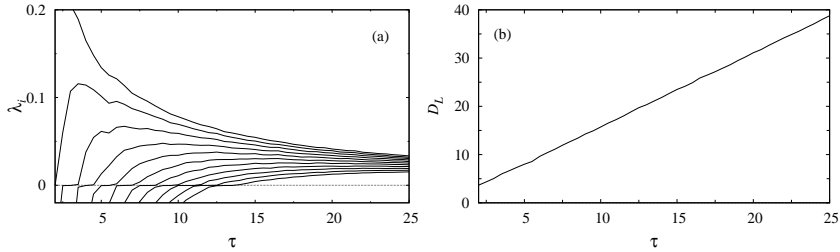}}
\caption{\label{chap2_fig8a}(a) Spectrum of the first eleven largest Lyapunov exponents for
the  values of the parameters $a=1, b=5$ in the range of delay time  $\tau\in(2,25)$
and (b) Kaplan-Yorke dimension $D_L$  in the corresponding range of
delay time for the Ikeda system~(\ref{eq1.01}).} 
\end{figure}

\section{The Ikeda time-delay system}

The Ikeda system was introduced to describe the dynamics of an optical bistable
resonator and it was shown that the transmitted light from a ring cavity
containing  a nonlinear dielectric medium undergoes a transition from a stationary
state to periodic and nonperiodic states, when the intensity of the incident
light is increased.  It has also been shown that the nonperiodic state is
characterized by a chaotic variation of the light intensity and associated
broadband noise in the power spectrum~\cite{kihd1980}. Ikeda system is well
known for delay induced chaotic behavior~\cite{kikm1987,cmdhz2001,mlber1987} and
this system has also been receiving focus on synchronization studies recently
~\cite{huv2000,emsss2002204,emsss2002pla,emsss2005}. 
The Ikeda model is
specified by the state equation
\begin{equation}
\dot x=-ax(t) -b\sin x(t-\tau),
\label{eq1.01}
\end{equation}
where $a>0$ and $b>0$ are the parameters and $\tau$ is the delay time.
Physically $x(t)$ is the phase lag of the electric field across the resonator
and thus may clearly assume both positive and negative values, $\alpha$ is the
relaxation coefficient, $b$ is the laser intensity injected into the system and
$\tau$ is the round-trip time of the light in the resonator. Typical chaotic
and hyperchaotic  chaotic attractors of the Ikeda system are shown in
Figs.~\ref{ikeda_att}a and b for values of delay times $\tau=2$ and $\tau=3$,
respectively, while the other parameter values are fixed as $a=1.0$
and  $b=5$. The first eleven largest Lyapunov exponents of the Ikeda system for
the parameters $a=1.0, b=5$ in the range of delay time $\tau \in (2,25)$ are
shown in  Fig.~\ref{chap2_fig8a}a and the corresponding Kaplan-Yorke Lyapunov
dimension calculated using the formula
\begin{equation}
D_L=j+\frac{\sum_{i=1}^{j}\lambda_i}{\left|\lambda_{j+1}\right|},
\label{lyadim}
\end{equation}
where j is the largest integer for which $\lambda_1 + ... + \lambda_j \ge 0$, is
shown in Fig.~\ref{chap2_fig8a}b.

\section{CPS in coupled Ikeda time-delay systems}
We consider the following unidirectionally coupled drive $x_1(t)$ and
response $x_2(t)$ systems
\begin{equation}
\dot{x}_1(t)=-ax_1(t)+b_1\sin x_1(t-\tau_1), \nonumber\\
\label{eq1.02}
\end{equation}
\begin{equation}
\dot{x}_2(t)=-ax_2(t)+b_2\sin x_2(t-\tau_2)+b_3\sin x_1(t-\tau_1),
\label{eq1.03}
\end{equation}
where the parameters are fixed as $a=1.0, b_1=b_2=5.0$. The delay times
$\tau_1=2$ and $\tau_2=3$ provide parameter mismatch between the drive,
$x_1(t)$, and the  response, $x_2(t)$, systems and $b_3$ is the coupling
strength. In the absence of the coupling both systems evolve independently
and the attractor of the drive system shown in Fig.~\ref{ikeda_att}a is chaotic
for the value of delay time $\tau=2$ and that of the response system shown in
Fig.~\ref{ikeda_att}b is hyperchaotic with two positive Lyapunov exponents for
the value of delay time $\tau=3$ as evidenced from the spectrum of Lyapunov
exponents shown in Fig.~\ref{chap2_fig8a}a. Hence both systems are
qualitatively different and their attractors shown in Figs.~\ref{ikeda_att} are
highly non-phase-coherent with interwoven trajectories exhibiting complex
topological properties. When the coupling strength $b_3$ is increased from zero,
the degree of chaotic phase synchronization between the drive and the response
systems increases after certain threshold value of the coupling strength and
finally they become phase synchronized fully. However, further increase in the
coupling strength does not lead to a transition to  generalized synchronization
even for appreciably larger value of $b_3$ unlike the case of coupled piecewise
linear delay systems or coupled Mackey-Glass
systems~\cite{dvskml2006,dvskml2008}. Now these results are depicted using
recurrence based indices, namely, $P(t)$, CPR, JPR and SPR.

\subsection{Recurrence based indices}

Synchronization transition in coupled Ikeda systems (\ref{eq1.02}) and
(\ref{eq1.03}), that is from desynchronized state to phase synchronized state
and then possibly to generalized synchronized state, can be analyzed by means of
recurrence based indices even when the corresponding attractors have complex
topological properties. The generalized autocorrelation function $P(t)$ has been
introduced in Refs.~\cite{nmmcr2007,mcrmt2005} as
\begin{equation}
P(t)=\frac{1}{N-t} \sum_{i=1}^{N-t} \Theta(\epsilon-||X_i-X_{i+t}|| ), 
\label{pbt}  
\end{equation}
where $\Theta$ is the Heaviside function, $X_i$ is the $i$th data  
corresponding to either the drive variable, $x_1$,  or the response variable,
$x_2$, and $\epsilon$ is a predefined threshold. $||.||$ is the  Euclidean norm
and $N$ is the number of data points. $P(t)$ can be considered as a statistical
measure about how often the phase $\phi$ has increased by $2\pi$ or multiples of
$2\pi$ within the time $t$ in the original space.  If two systems are in CPS,
their phases increase on average by $K\cdot 2\pi$, where $K$ is a natural number,
within the same time interval $t$. The value of $K$ corresponds to the number of
cycles when $||X(t+T)-X(t)||\sim 0,$ or equivalently when $||X(t+T)-X(t)|| <
\epsilon$, where $T$ is the period of the system. Hence, looking at the
coincidence of the positions of the maxima of $P(t)$ for both the systems
(\ref{eq1.02}) and (\ref{eq1.03}), one can
qualitatively identify CPS.

\begin{figure}
\centering
\resizebox{0.6\columnwidth}{!}{
\includegraphics{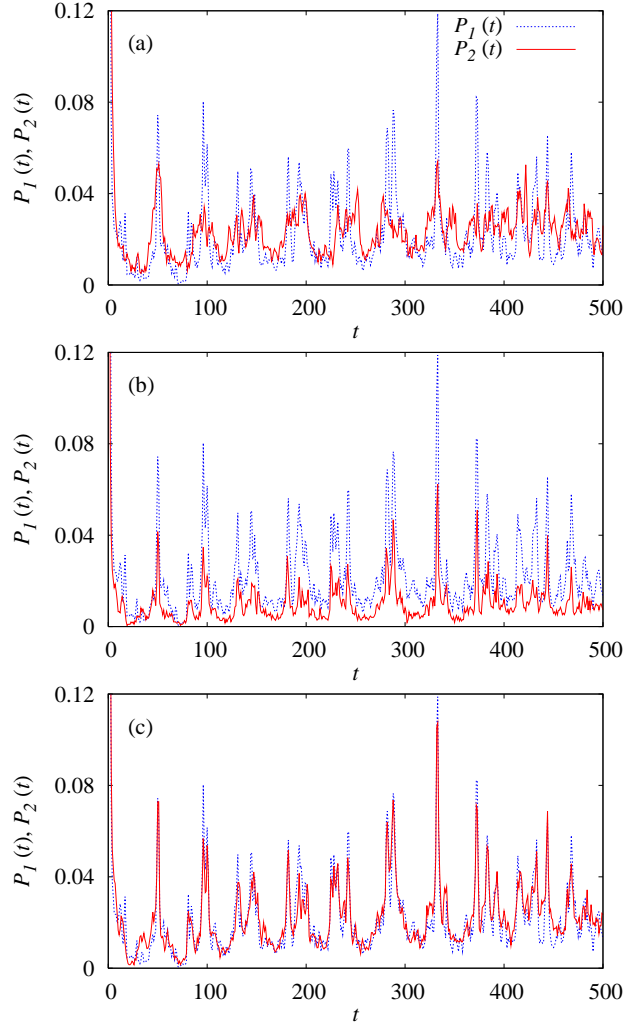}}
\caption{\label{chap8_fig39}(Color online) Generalized autocorrelation functions
of both the drive $P_1(t)$ and the response $P_2(t)$ systems. (a)
Non-synchronization for $b_3=4$, (b) Approximate phase synchronization for
$b_3=10$ and (c) Phase synchronization for $b_3=20$.}
\end{figure}

A criterion to quantify CPS is the cross correlation 
coefficient between the drive, $P_1(t)$, and the response,
$P_2(t)$, which can be defined as Correlation of Probability of Recurrence
(CPR),
\begin{equation}
CPR=\langle \bar{P_1}(t)\bar{P_2}(t)\rangle/\sigma_1\sigma_2,
\end{equation}
where $\bar{P}_{1,2}$ means that the mean value has been subtracted and
$\sigma_{1,2}$ are the standard deviations of $P_1(t)$ and $P_2(t)$,
respectively.  If the two systems (\ref{eq1.02}) and (\ref{eq1.03}) are in CPS,
the probability of recurrence is maximal at the same time $t$ and CPR $\approx
1$. If they are not in CPS, the maxima do not occur simultaneously and hence one
can expect a drift in the probability of recurrences which results in low
values of  CPR.

When  the coupled Ikeda systems (\ref{eq1.02}) and (\ref{eq1.03}) are in
generalized synchronization, two close states in the phase space of the drive
variable correspond to that of the response.  Hence the neighborhood identity is
preserved in phase space.  Since the recurrence plots are nothing but a record
of the neighborhood of each point in the phase space, one can expect that their
respective recurrence plots are almost identical.  Based on these facts the
following two indices can be calculated as proposed in ~\cite{mcrmt2005} to
quantify GS for the Ikeda systems similar to the coupled piecewise linear and
Mackey-Glass systems analyzed by us recently~\cite{dvskml2006,dvskml2008}.

First, the authors of \cite{mcrmt2005} proposed the 
 Joint Probability of Recurrences (JPR),
\begin{equation}
JPR=\frac{\frac{1}{N^2} \sum_{i,j}^N \Theta(\epsilon_x-||X_i-X_j||
)\Theta(\epsilon_y-||Y_i-Y_j||)-RR}{1-RR}  
\label{jpr}  
\end{equation}
where $RR$ is rate of recurrence, $\epsilon_x$ and $\epsilon_y$ are thresholds
corresponding to the drive and response systems, respectively such that
$RR_X=RR_Y=RR$ and $X_i$ is the $i$th data corresponding to the drive variable
$x_1$ and $Y_i$ is the $i$th data corresponding to the response variable $x_2$.
RR measures the density of recurrence points and it is fixed as 0.02
\cite{mcrmt2005}. JPR is close to $1$ for systems in GS and is small when they
are not  in GS. The second index depends on the coincidence of the  probability
of recurrence, which is defined as Similarity of Probability of Recurrence
(SPR),
\begin{equation}
SPR=1-\langle(\bar{P_1}(t)-\bar{P_2}(t))^2\rangle/\sigma_1\sigma_2.
\end{equation}  
SPR is again of order $1$ if the two systems are in GS and approximately zero or
negative if they evolve independently.
\begin{figure}
\centering
\resizebox{0.7\columnwidth}{!}{
\includegraphics{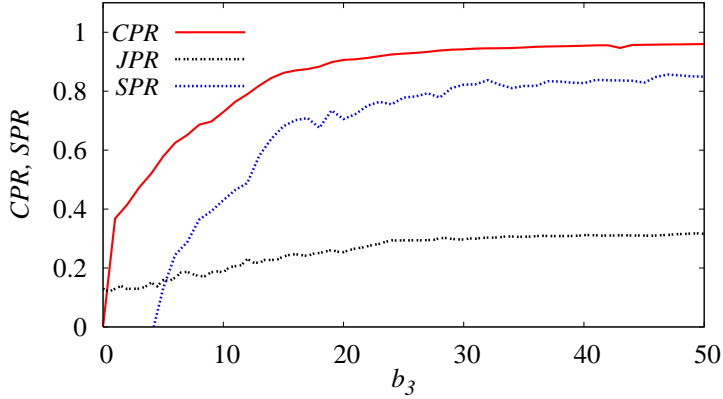}}
\caption{\label{ikeda_spec_cpr}(Color online) Indices CPR, SPR as a function of
coupling strength $b_3\in(0,50)$.}
\end{figure}

Now, we will apply these concepts to the original non-phase-coherent attractors
shown in Figs.~\ref{ikeda_att}, when coupling is introduced as in
Eqs.~(\ref{eq1.02}) and (\ref{eq1.03}). We  estimate these recurrence based
measures from $5000$ data points after barring out sufficient transients with
the integration step $h=0.01$ and sampling rate $\Delta t =100$. The generalized
autocorrelation functions $P_1(t)$ of the drive $x_1(t)$ system and $P_2(t)$ of
the response $x_2(t)$ systems are depicted in Figs.~\ref{chap8_fig39} for
different values of the coupling strength.  The maxima of the generalized
autocorrelation functions $P_1(t)$ and $P_2(t)$  do not occur simultaneously
(Fig.~\ref{chap8_fig39}a) and there exists a drift between them for the value of
the coupling strength $b_3=4$ and hence both the systems evolve independently. 
This fact is also reflected in the rather low values of the indices CPR, JPR and
SPR as shown in Fig.~\ref{ikeda_spec_cpr}. Looking into the details of the
generalized autocorrelation functions in Fig.~\ref{chap8_fig39}b for the
value of the coupling strength $b_3=10$, we find that the main oscillatory
dynamics becomes locked and hence the large amplitude peaks (maxima) of $P_1(t)$
and $P_2(t)$ coincide while small amplitude peaks do not. This behavior is
observed in the range of $b_3\in(4.2,20)$, which corresponds to the transition
regime (approximate CPS) and this is also indicated by a smooth increase in the
value of CPR (Fig.~\ref{ikeda_spec_cpr}) towards the value $1$.

\begin{figure}
\centering
\resizebox{0.7\columnwidth}{!}{
\includegraphics{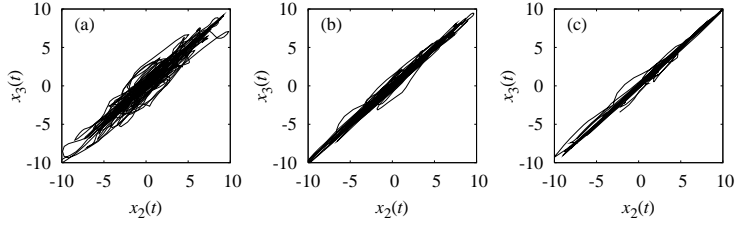}}
\caption{\label{ikeda_gs} The plot of response variable $x_2(t)$, vs, auxiliary
variable  $x_3(t)$ for the values of the coupling strengths (a) $b_3=30$, (b)
$b_3=40$  and (c) $b_3=50$.}
\end{figure}

Further increase in the value of the coupling strength beyond $b_3=20$ results
in an almost perfect locking of all the oscillatory dynamics of the coupled
system. Consequently, a majority of the positions of the peaks in the generalized
autocorrelation functions  $P_1(t)$ and $P_2(t)$ agree with each other as
illustrated in Fig.~\ref{chap8_fig39}c for the value of $b_3=20$. However, it is
observed that  the magnitude of  peaks are generally of  different values and
this difference in the heights of peaks indicates that there is no correlation
in the amplitudes of both the systems. This is in accordance with strongly
bounded nature of phase difference and further increase in the value of the
coupling strength results in a saturation in the value of CPR $\approx 1$  as
seen from the Fig.~\ref{ikeda_spec_cpr}, which is  a strong indication for the
existence of CPS. 

\begin{figure}
\centering
\resizebox{0.7\columnwidth}{!}{
\includegraphics{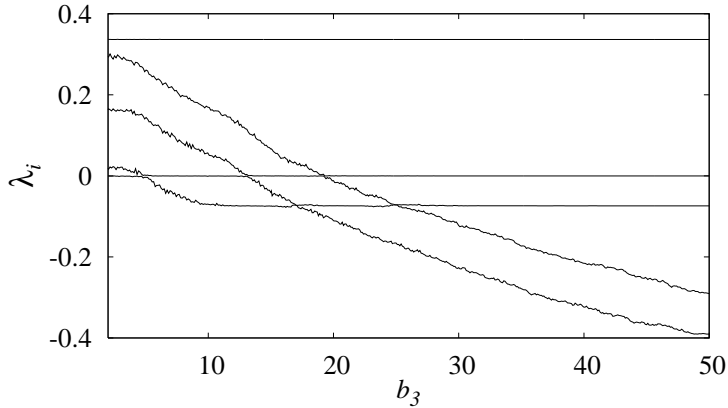}}
\caption{\label{ikeda_cou_lya} Spectrum of first five largest Lyapunov exponents
of the coupled Ikeda systems (\ref{eq1.02}) as a function of
coupling strength $b_3\in(0,50)$.}
\end{figure}

Even for a very large value of the coupling strength, say $b_3=50$, the
amplitudes of the maxima in the generalized autocorrelation function do not
coincide and hence one does not find an indication towards the exact  GS. This
is further confirmed from the rather lower values of JPR and SPR depicted in
Fig.~\ref{ikeda_spec_cpr}. This scenario is in contrast to our earlier
studies~\cite{dvskml2006,dvskml2008} where there exists transition from phase to
generalized synchronization within reasonable range of values of the coupling
strength in the case of piece-wise linear and Mackey-Glass time-delay systems.
This is further confirmed from the auxiliary system approach by augmenting the
coupled Ikeda systems ((\ref{eq1.02}) and (\ref{eq1.03})) with an additional
auxiliary system for the variable $x_3(t)$ identical to the response system,
satisfying the equation
\begin{equation}
\dot{x}_3(t)=-ax_3(t)+b_2\sin x_3(t-\tau_2)+b_3\sin x_1(t-\tau_1).
\label{eq_aux}
\end{equation}
We have analyzed numerically the combined system of equations (\ref{eq1.02}), 
(\ref{eq1.03}) and (\ref{eq_aux}). The plot of the response variable  $x_2(t)$
vs the auxiliary variable $x_3(t)$ for the values of the coupling strengths
$b_3=30, 40$ and $50$ are depicted in Figs.~\ref{ikeda_gs}a, \ref{ikeda_gs}b and
\ref{ikeda_gs}c, respectively. As may be noted that for none of these values one
obtains a sharp diagonal line, indicating only the existence of approximate GS.
A possible reason for this is that the largest Lyapunov exponents of the
response system do not attain negative saturation even for larger values of the
coupling strength as shown in Fig.~\ref{ikeda_cou_lya}. Hence  there does not
seem to exist exact generalized synchronization between the coupled Ikeda
systems in the explored range of parameters by us, even though CPS does exist in
this range.

\subsection{Spectrum of Lyapunov exponents}

The transition from non-synchronization to CPS is also characterized by changes
in the spectrum of  Lyapunov exponents of the coupled time-delay systems
(\ref{eq1.02}) and (\ref{eq1.03}). The spectrum of the  first five largest
Lyapunov exponents of the coupled Ikeda systems is shown in
Fig.~\ref{ikeda_cou_lya}. The null Lyapunov exponent of the response system
$x_2(t)$ becomes negative at the value of the coupling strength $b_3=4.2$ while
the other two largest Lyapunov exponents remain positive, which is a typical
characteristic feature for the onset of CPS in the coupled systems. The second
least positive Lyapunov exponent becomes negative at the value of the coupling
strength $b_3=13.5$, an indication of onset of correlation in amplitudes of both
the interacting dynamical systems, while the largest positive Lyapunov exponent
of the response  system becomes negative at the value of $b_3=20$.  This is a
strong indication that in this rather complex attractor the amplitudes become
somewhat interrelated already at the transition to PS in the range
$b_3\in(4.2,20)$ in agreement with our earlier results in coupled Mackey-Glass
systems and  (as in the  funnel attractor~\cite{gvobh2003}). However, it is
observed that there does not exist generalized synchronization between the
coupled Ikeda systems even for larger values of the  coupling strength in
contrast to our earlier studies~\cite{dvskml2006,dvskml2008} and this is evident
from the value of the Lyapunov exponents of the response system becoming
increasingly negative without attaining saturation.

\subsection{Concept of Localized Sets}

Recently, an interesting framework for identifying phase synchronization without
having explicitly the measure of the phase, namely the concept of localized
sets, has been introduced~\cite{tpmsb2007}. The basic idea of this concept is
that one has to define a typical event in one of the coupled oscillators and
then observe the other oscillator whenever this event occurs.  These
observations give rise to a set $D$. Depending upon the property of this set $D$
one can state whether there PS exists or not. The coupled oscillators evolve
independently if the sets obtained by observing the corresponding events in both
the oscillators spread over the attractors of the oscillators. On the other
hand, if the sets are localized on the attractors then PS exist between the
interacting oscillators. 

\begin{figure}
\centering
\resizebox{0.7\columnwidth}{!}{
\includegraphics{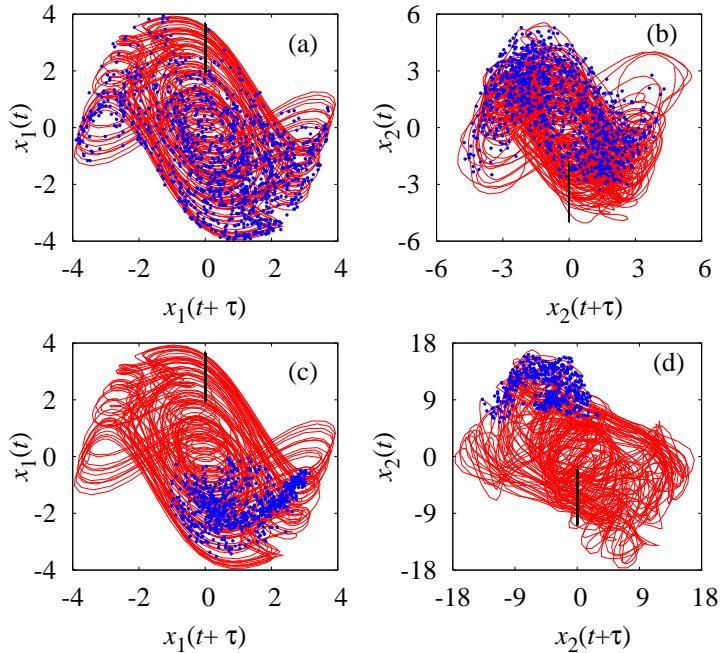}}
\caption{\label{ikeda_ls} (Color online) (a) and (c) attractor of the drive
system, (b) and (d) attractor of the response system. The bars indicate the
events in the corresponding attractors. In (a) and (b) the sets spread over the
attractor  and hence there is no CPS for the value of the coupling strength
$b_3=4.0$ and, in (c) and (d) the sets are localized confirming the existence of
CPS for $b_3=20.0$.}
\end{figure}

We have confirmed the existence of CPS in the coupled Ikeda time-delay systems
also by using the concept of localized set. We have defined the event in the
attractor of the drive system as a segment characterized by $x_1(t+\tau)=0$ and
$x_1(t)>2.0$ and another event in the response system as a segment characterized
by $x_2(t+\tau)=0$ and $x_2(t)<-3.0$, which are shown as black lines in
Fig.~\ref{ikeda_ls}. The sets obtained by observing the response Ikeda system
whenever the defined event occurs in the drive system and vice versa are shown
as dots in Figs.~\ref{ikeda_ls}a and~\ref{ikeda_ls}b, respectively,  for the
value of the coupling strength $b_3=4.0$, for which there is no CPS as discussed
earlier and hence the sets are spread over the attractors.  On the other hand
for the value of the coupling strength $b_3=20$ for which CPS exists as seen
from Figs.~\ref{chap8_fig39}-\ref{ikeda_cou_lya}, the sets are localized as
shown in Figs.~\ref{ikeda_ls}c and \ref{ikeda_ls}d confirming the existence of
CPS in the coupled Ikeda systems.

\section{Summary and Conclusion}

We have identified the existence of CPS in coupled Ikeda time-delay systems
which possess highly non-phase-coherent chaotic and hyperchaotic attractors with
complex topology.  In particular, we have shown that there is a typical
transition from a nonsynchronized state to CPS as a function of the coupling
strength.  We have characterized this transition in terms of recurrence based
indices such as $P(t)$, CPR, JPR and SPR, and quantified the different
synchronization regimes in terms of them.  The transition is also confirmed by
the typical transition in the Lyapunov exponents of the coupled Ikeda time-delay
systems. Further, we have also confirmed the existence of CPS using the concept
of localized sets. We have found that the recurrence based techniques are more
efficient than the other conventional techniques available in the literature to
identify CPS in higher dimensional systems, in particular in time-delay systems.
It is also of interest to find out a suitable general transformation to include
the attractors of large class of time-delay systems, including coupled Ikeda
systems, which transforms the non-phase-coherent attractors into smeared limit
cycle like attractors. Work is in progress on this aspect.

\begin{acknowledgement}
The work of D. V. S. and M. L. has been supported by a Department of Science and
Technology, Government of India sponsored research project. The work of M. L. is
supported by a DST Ramanna Fellowship. J. K has been supported by his Humboldt-CSIR research award and  NoE BIOSIM (EU)
Contract No. LSHB-CT-2004-005137. 
\end{acknowledgement}

\end{document}